\documentclass[12pt]{article}
\usepackage{epsfig}

\newcommand\pubnumber{DESY-01-001}
\newcommand\pubdate{January 2001}
\newcommand\hepnumber{hep-ph/0101080}

\def\csumb{Deutsches Elektronen-Synchroton DESY, 22603 Hamburg, Germany}

\def\Title#1{\begin{center} {\Large\bf #1 } \end{center}}
\def\Author#1{\begin{center}{ \sc #1} \end{center}}
\def\Address#1{\begin{center}{ \it #1} \end{center}}

\newcommand\pubblock{\rightline{\begin{tabular}{l} \pubnumber\\
         \pubdate\\ \hepnumber \end{tabular}}}
\newenvironment{Abstract}{\begin{quotation}  }{\end{quotation}}
\newenvironment{Presented}{\begin{quotation} \begin{center} 
             Presented at the\end{center}
      \begin{center}\begin{large}}{\end{large}\end{center} \end{quotation}}
\def\Acknowledgments{\bigskip  \bigskip \begin{center}
          \large\bf Acknowledgments\end{center}}

\makeatletter
\def\section{\@startsection{section}{0}{\z@}{5.5ex plus .5ex minus
 1.5ex}{2.3ex plus .2ex}{\large\bf}}
\def\subsection{\@startsection{subsection}{1}{\z@}{3.5ex plus .5ex minus
 1.5ex}{1.3ex plus .2ex}{\normalsize\bf}}
\def\subsubsection{\@startsection{subsubsection}{2}{\z@}{-3.5ex plus
-1ex minus  -.2ex}{2.3ex plus .2ex}{\normalsize\sl}}

\renewcommand{\@makecaption}[2]{%
   \vskip 10pt
   \setbox\@tempboxa\hbox{\small #1: #2}
   \ifdim \wd\@tempboxa >\hsize     
       \small #1: #2\par          
     \else                        
       \hbox to\hsize{\hfil\box\@tempboxa\hfil}
   \fi}

 \def\citenum#1{{\def\@cite##1##2{##1}\cite{#1}}}
 
\newcount\@tempcntc
\def\@citex[#1]#2{\if@filesw\immediate\write\@auxout{\string\citation{#2}}\fi
  \@tempcnta\z@\@tempcntb\m@ne\def\@citea{}\@cite{\@for\@citeb:=#2\do
    {\@ifundefined
       {b@\@citeb}{\@citeo\@tempcntb\m@ne\@citea\def\@citea{,}{\bf ?}\@warning
       {Citation `\@citeb' on page \thepage \space undefined}}%
    {\setbox\z@\hbox{\global\@tempcntc0\csname b@\@citeb\endcsname\relax}%
     \ifnum\@tempcntc=\z@ \@citeo\@tempcntb\m@ne
       \@citea\def\@citea{,}\hbox{\csname b@\@citeb\endcsname}%
     \else
      \advance\@tempcntb\@ne
      \ifnum\@tempcntb=\@tempcntc
      \else\advance\@tempcntb\m@ne\@citeo
      \@tempcnta\@tempcntc\@tempcntb\@tempcntc\fi\fi}}\@citeo}{#1}}
\def\@citeo{\ifnum\@tempcnta>\@tempcntb\else\@citea\def\@citea{,}%
  \ifnum\@tempcnta=\@tempcntb\the\@tempcnta\else
  {\advance\@tempcnta\@ne\ifnum\@tempcnta=\@tempcntb \else\def\@citea{--}\fi
    \advance\@tempcnta\m@ne\the\@tempcnta\@citea\the\@tempcntb}\fi\fi}
\makeatother

\def\beq{\begin{equation}}
\def\eeq#1{\label{#1}\end{equation}}
\def\eeqn{\end{equation}}

\newenvironment{Eqnarray}%
   {\arraycolsep 0.14em\begin{eqnarray}}{\end{eqnarray}}
\def\beqa{\begin{Eqnarray}}
\def\eeqa#1{\label{#1}\end{Eqnarray}}
\def\eeqan{\end{Eqnarray}}

\let\bar=\overbar

\def\etal{{\it et al.}}
\def\ie{{\it i.e.}}

\def\Dslash{\not{\hbox{\kern-4pt $D$}}}
\def\dslash{\not{\hbox{\kern-2pt $\del$}}}

\def\msb{{\bar{\ssstyle M \kern -1pt S}}}

\def\lsim{\mathrel{\raise.3ex\hbox{$<$\kern-.75em\lower1ex\hbox{$\sim$}}}}
\def\gsim{\mathrel{\raise.3ex\hbox{$>$\kern-.75em\lower1ex\hbox{$\sim$}}}}

\begin{document}
\begin{titlepage}
\pubblock

\vfill
\def\thefootnote{\fnsymbol{footnote}}
\Title{Exclusive QCD}
\vfill
\Author{Markus Diehl}
\Address{\csumb}
\vfill
\begin{Abstract}
I give a brief introduction to the physics of generalized parton
distributions and distribution amplitudes. I then report on the status
of the calculation of radiative corrections for the exclusive
processes where these quantities occur.
\end{Abstract}
\vfill
\begin{Presented}
5th International Symposium on Radiative Corrections \\ 
(RADCOR--2000) \\[4pt]
Carmel CA, USA, 11--15 September, 2000
\end{Presented}
\vfill
\end{titlepage}
\def\thefootnote{\arabic{footnote}}
\setcounter{footnote}{0}

\section{Introduction}

In recent years a formalism has been developed which highlights the
close connection between exclusive and inclusive strong interaction
processes. The cornerstones of this formalism are the concepts of
generalized parton distributions and generalized distribution
amplitudes. These quantities contain valuable information on the
non-perturbative transition between partons and hadrons, whose
understanding remains one of the great outstanding tasks in QCD. They
can be accessed in several exclusive reactions that are within the
reach of current and planned experimental facilities. In this
contribution I will first give a brief overview of the formalism and
the physics behind it, and then report on the status of the
calculation of radiative corrections in this context.

\begin{figure}[b!]
\begin{center}
\epsfig{file=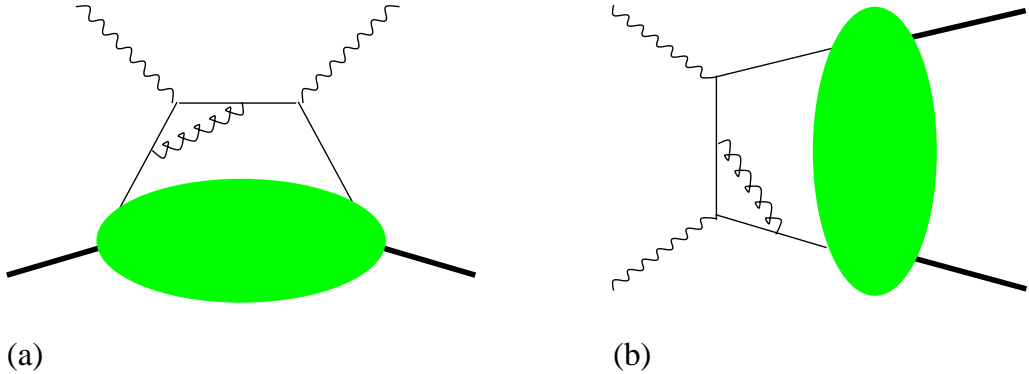,width=0.9\textwidth}
\caption[0]{\label{fig:compton} (a)~A diagram for the Compton
amplitude $\gamma^* p \to \gamma^* p$ in Bjorken kinematics. (b)~A
diagram for the corresponding amplitude $\gamma^*\gamma^*\to p\bar{p}$
in the crossed channel.}
\end{center}
\end{figure}

A key process in the development of the QCD improved parton model has
been inclusive deep inelastic scattering, $e p \to e X$, which via the
optical theorem is conveniently expressed in terms of the imaginary
part of the forward Compton amplitude, $\gamma^*(q) + p(p) \to
\gamma^*(q) + p(p)$. In the Bjorken region of large photon virtuality
$Q^2=-q^2$ and c.m.\ energy $(p+q)^2$, this amplitude factorizes into
a perturbatively calculable scattering process at the level of quarks
and gluons and process independent matrix elements
\begin{equation}
  \label{pdf}
\langle\, p(p) |\, {\cal O}(\lambda) \,| p(p) \rangle .
\end{equation}
Here ${\cal O}(\lambda)$ stands for quark or gluon operators
$\bar{\psi}(0)\, n_\mu\gamma^\mu \psi(\lambda n)$, $\bar{\psi}(0)\,
n_\mu\gamma^\mu\gamma_5 \psi(\lambda n)$, $n_\mu n_\nu\,
F^{\mu\alpha}(0) F^{\nu}{}_{\alpha}(\lambda n)$, $n_\mu n_\nu\,
F^{\mu\alpha}(0) \widetilde{F}^{\nu}{}_{\alpha}(\lambda n)$, whose
fields are separated by a light-like distance $\lambda n$ (\ie,
$n^2=0$). These matrix elements, represented by a blob in
fig.~\ref{fig:compton}a, are parameterized by parton distributions;
they describe the transition between hadronic and partonic degrees of
freedom.

This factorization into short- and long-distance dynamics is actually
more general. It also holds for the nonforward Compton amplitude,
$\gamma^*(q) + p(p) \to \gamma^*(q') + p(p')$, in a generalization of
Bjorken kinematics, namely if the c.m.\ energy $(p+q)^2$ and at least
one of the photon virtualities $|q^2|$, $|q'^2|$ are large while the
invariant momentum transfer $(p-p')^2$ is small
\cite{Muller:1994fv,Ji:1997ek,Radyushkin:1997ki}. In the particular
case where the outgoing photon is on shell, one speaks of deeply
virtual Compton scattering, which can be accessed in the physical
process $e p \to e p \gamma$, \ie, in exclusive electroproduction of a
photon. The non-perturbative physics is now described by matrix
elements with the \emph{same} operators as before, but between
\emph{different} hadron states,
\begin{equation}
  \label{skewed}
\langle p(p') |\, {\cal O}(\lambda) \,| p(p) \rangle .
\end{equation}
The nonzero momentum transfer to the proton implies that the momenta
of the two parton lines attached to the blob in
fig.~\ref{fig:compton}a must differ as well. A simple calculation
shows that in particular their momentum fractions with respect to the
hadrons cannot be equal. For this reason, the generalized parton
distributions which parameterize the matrix elements (\ref{skewed})
are often called ``skewed''.

A completely analogous type of factorization occurs in the crossed
channel, \ie, in $\gamma^*(q) + \gamma^*(q') \to p(p) + \bar{p}(p')$,
if at least one of the photon virtualities is large, in particular
compared with the invariant mass $(p+p')^2$ of the produced hadron
pair \cite{Muller:1994fv,Diehl:1998dk}. A corresponding diagram is
shown in fig.~\ref{fig:compton}b. Matrix elements
\begin{equation}
  \label{gda}
\langle p(p)\, \bar{p}(p')|\, {\cal O}(\lambda) \,| 0 \rangle
\end{equation}
with again the same operators as before now parameterize the
non-perturbative transition from a quark-antiquark or gluon pair to
the final state hadrons. In addition to the $p\bar{p}$ system one can
consider a wide range of hadrons, say $\pi\pi$ or $K K$, which are not
easily available as beam particles in Compton scattering. The
production mechanism represented in fig.~\ref{fig:compton}b is the
same as in the process $\gamma^* \gamma \to \pi$, where the
nonperturbative input is represented by the quark-antiquark
distribution amplitude of the pion. Data on this process have in fact
provided one of the best available constraints so far on this
important quantity \cite{Gronberg:1998fj}. The matrix elements
(\ref{gda}) are a direct generalization of usual distribution
amplitudes, where $\langle p \bar{p}|$ is replaced by a single-meson
state.

\section{Some physics aspects}

\begin{figure}
\begin{center}
\epsfig{file=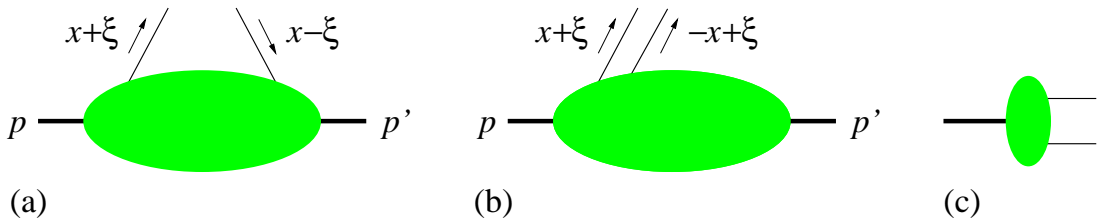,width=0.95\textwidth}
\caption[0]{\label{fig:regimes} (a)~The kinematical regime where a
generalized parton distribution describes the emission and
reabsorption of a parton.  (b)~The regime where it describes the
emission of a quark-antiquark or gluon pair. This is reminiscent of a
meson distribution amplitude, shown in (c).}
\end{center}
\end{figure}

As a consequence of the finite momentum transfer to the proton,
generalized parton distributions admit two different kinematical
regimes. In the first, they describe the emission of a parton with
momentum fraction $x+\xi$ from the parent hadron and its reabsorption
with momentum fraction $x-\xi$, see fig.~\ref{fig:regimes}a. In the
limit where $p=p'$ one has $\xi=0$ and recovers the usual parton
distributions. In a second regime, which does not exist for $p=p'$,
one has the coherent emission of a quark-antiquark or gluon pair from
the parent hadron of momentum $p$, leaving the hadron with momentum
$p'$, see fig.~\ref{fig:regimes}b. One is thus sensitive to aspects of
the proton structure that cannot be accessed by the ordinary parton
densities. The second regime is reminiscent of a distribution
amplitude, shown in fig.~\ref{fig:regimes}c, where there is no hadron
left behind after emission of the partons. We will encounter an
important manifestation of this similarity in
section~\ref{sec:evolution}.

\begin{figure}[t!]
\begin{center}
\epsfig{file=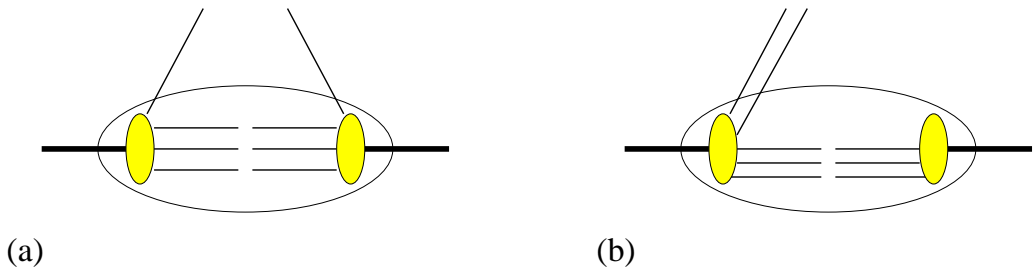,width=0.9\textwidth}
\caption[0]{\label{fig:overlap} Representation of a parton
distribution in terms of light-cone wave functions, denoted by small
blobs, in the two regimes of fig.~\protect\ref{fig:regimes}a and
b. Notice that the second regime involves wave functions with
different parton content.}
\end{center}
\end{figure}

It has long been known that the usual parton distributions can be
represented in terms of light-cone wave functions, which completely
specify the structure of a hadron in terms of quark and gluon
configurations \cite{Brodsky:1989pv}. In this representation, depicted
in fig.~\ref{fig:overlap}a, the wave functions appear \emph{squared},
which reflects the crucial parton model feature that parton densities
can be understood as classical probabilities for the emission of a
parton from a hadron. The wave function representation of generalized
parton distributions \cite{Brodsky:2000xy} provides a key to their
physical interpretation: they are not probabilities but
\emph{interference} terms of wave functions for different parton
configurations in a hadron. In this sense they contain characteristic
information on the quantum fluctuations of QCD bounds states, going
beyond the classical probability picture of the parton model.

Generalized parton distributions have a rich structure in spin, since
the helicities of the partons and the hadrons can be varied
independently. Of particular interest are those combinations where the
helicity difference on the hadron side is not compensated on the
parton side: in that case angular momentum conservation must be
ensured by a transfer of orbital angular momentum, which is possible
if there is a finite transfer of transverse momentum. Thus,
generalized distributions carry information on the orbital angular
momentum of partons---information that is hard to access otherwise. If
one takes moments of the generalized distributions in the momentum
fraction $x$, the operators in the matrix elements (\ref{skewed}) are
transformed into local operators, \ie, one obtains form factors of
various local currents. A sum rule due to Ji \cite{Ji:1997ek} states
that the second moment of a particular combination of generalized
distributions gives a form factor whose value at zero momentum
transfer is the \emph{total} angular momentum of quarks in the proton,
consisting both of a spin and an orbital part.

The generalized distribution amplitudes of fig.~\ref{fig:compton}b are
intimately connected with generalized parton distributions by crossing
symmetry. Their moments are in fact related by an analytic
continuation in the Mandelstam invariant, \ie, $t=(p-p')^2$ for
distributions and $s=(p+p')^2$ for distribution amplitudes. On the
other hand, generalized distribution amplitudes contain physics quite
distinct from that of parton distributions: they involve not only the
partonic structure of a single hadron, but also the interactions
between hadrons. They parameterize what one might call the ``exclusive
limit'' of fragmentation, \ie, of the transition between parton and
hadron degrees of freedom. It is interesting to note that one can make
a connection with phenomenologically successful pictures of
hadronization such as the Lund string model \cite{Maul:2001ky}.

\section{Processes}

\begin{figure}[t!]
\begin{center}
\epsfig{file=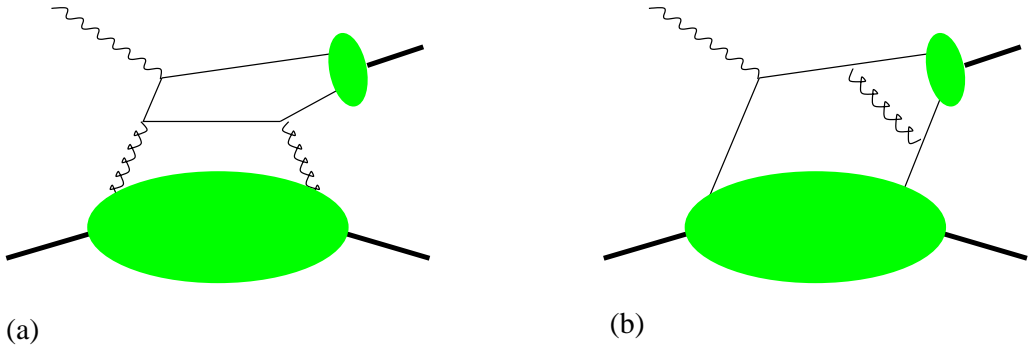,width=0.9\textwidth}
\caption[0]{\label{fig:mesons} Born level diagrams for the exclusive
production of a meson from a deeply virtual photon. The large blobs
denote generalized gluon or quark distributions, and the small blobs
the meson distribution amplitude. Diagram (a) only contributes for
mesons with negative charge conjugation parity.}
\end{center}
\end{figure}

As we saw in the introduction, generalized parton distributions can be
accessed in deeply virtual Compton scattering, measurable by
electroproduction $e p\to e p \gamma$. Another class of processes
where they occur is exclusive electroproduction of a meson instead of
a photon, $e p\to e p \rho^0$, $e p\to e n \pi^+$, etc. Example
diagrams are shown in fig.~\ref{fig:mesons}. Notice that for vector
mesons both quark and gluon distributions contribute at leading order
in $\alpha_s$. This is in contrast to Compton scattering, where gluons
only appear at the level of one-loop corrections, as they do in
inclusive deep inelastic scattering.

To date, we have no experimental determinations of generalized parton
distributions. However, first measurements of the above processes in
the relevant kinematical domain have been performed, in particular by
HERMES, H1, and ZEUS at DESY. Further and more precise investigations
are planned Jefferson Lab and at CERN (COMPASS), and several future
accelerator projects would be well suited for in-depth studies of
these processes.

Generalized distribution amplitudes can be accessed in
$\gamma^*\gamma^*$ or $\gamma^*\gamma$ processes at $e^+e^-$
colliders, and cross section estimates \cite{Diehl:2000uv} indicate
that the production of pion pairs $\gamma^*\gamma\to \pi\pi$ should be
well in the reach of BABAR, BELLE, and CLEO.

\section{Radiative Corrections}

As in all applications of QCD factorization, radiative corrections in
our context manifest themselves in two ways. The process independent
hadronic matrix elements depend on a factorization scale via evolution
equations, whose kernels can be calculated in perturbation theory. On
the other hand, there are radiative corrections to the hard scattering
subprocesses for each individual reaction.

\subsection{Evolution} 
\label{sec:evolution}

The evolution of generalized distribution amplitudes is exactly the
same as the one of the usual distribution amplitudes for mesons,
described by the ERBL equations \cite{Lepage:1979zb}. This is not
surprising, since evolution arises from the renormalization of the
bilocal operators ${\cal O}(\lambda)$ and is insensitive to whether
the matrix elements~(\ref{gda}) involve a single meson or a
two-particle state with the same quantum numbers.

The evolution of general parton distributions on the other hand, is
more complex and of considerable theoretical interest. It is in fact
this aspect on which the earliest studies of these quantities have
focused on \cite{Muller:1994fv,Blumlein:1997pi}. In the regime of
fig.~\ref{fig:regimes}a, the evolution is similar to the standard
DGLAP evolution of parton densities, with a kernel that depends on the
parameter $\xi$ describing the longitudinal momentum transfer to the
partons (see fig.~\ref{fig:regimes}). In the regime of
fig.~\ref{fig:regimes}b, evolution acts in a similar way as ERBL
evolution, which highlights the close analogy between
figs.~\ref{fig:regimes}b and c.  The evolution equations for ordinary
parton distributions and for distribution amplitudes, although acting
in quite different ways, are thus intimately related, which stems from
the fact that these quantities are defined through the same bilocal
operators ${\cal O}(\lambda)$. The evolution kernels for generalized
parton distributions, called ``extended ERBL kernels'', contain the
usual DGLAP and ERBL kernels as limiting cases; in this sense the
evolution of generalized parton distributions \emph{interpolates}
between the two extremes of DGLAP and ERBL evolution.

The extended ERBL kernels have been calculated to LO by many
groups. They can be found to NLO accuracy in \cite{Belitsky:2000hf},
where conformal and supersymmetric constraints were employed in order
to reconstruct them from the known NLO DGLAP kernels. A numerical
study (limited to parton helicity independent distributions) showed
that the effect of NLO evolution was moderate compared with LO
evolution \cite{Belitsky:1999uk}. For the model distributions used
there, the difference between NLO and LO evolution was a few percent
for non-singlet distributions and not more than 10 to 30\% in the
singlet sector.

\subsection{The two-photon processes}

The one-loop corrections to deeply virtual Compton scattering have
been independently calculated by three groups
\cite{Ji:1998xh,Belitsky:2000sg}. In addition to diagrams like the one
in fig.~\ref{fig:compton}a they involve diagrams with the generalized
gluon distributions in the proton. In \cite{Ji:1998xh} one finds the
NLO results for the general nonforward amplitude $\gamma^*(q) + p(p)
\to \gamma^*(q') + p(p')$; in the limit $q=q'$ their imaginary parts
reduce to the well-known expressions for unpolarized and polarized
deep inelastic scattering. In \cite{Belitsky:2000sg} a numerical study
for $\gamma^* p \to \gamma p$ was performed, making an ansatz for the
yet unknown generalized quark and gluon distributions. It was found
that the NLO corrections can be large, up to about 50\%, and depend
sensitively on the value of the Bjorken variable $x_B = Q^2 /(2p\cdot
q)$.

By an analytic continuation of the hard scattering kernel, the
one-loop corrections for $\gamma^* \gamma \to \pi\pi$ have been
obtained from those for the general nonforward Compton amplitude
\cite{Kivel:1999sd}. Numerical studies show that the size of the
corrections is very sensitive to the relative size of the two-gluon
and the quark-antiquark distribution amplitudes
\cite{Kivel:1999sd,Kivel:2000rq}. In other words, this process may
offer an interesting way to investigate how strongly the two-pion
system couples to $gg$ in comparison with $q\bar{q}$.

\subsection{Power corrections}

The factorized description discussed so far is valid in the limit of
infinitely large photon virtuality $Q^2$, and at finite $Q^2$ there
are as usual corrections suppressed by powers of $1/Q$, up to
logarithmic terms.

An estimation of $1/Q^2$ corrections to deeply virtual Compton
scattering and to pion electroproduction has been made in
\cite{Vanttinen:1998pp} with the help of the renormalon technique,
resumming vacuum polarization insertions in the gluon lines of
figs.~\ref{fig:compton}a and \ref{fig:mesons}b. The corrections,
evaluated at $Q^2=4$~GeV$^2$, were found to grow with $x_B$ and to be
important (of order 10 to 50\%) for the Compton process. For pion
production they came out substantially larger (100\% and more), with a
strong dependence on the ansatz made for the generalized quark
distributions.

The structure of the $1/Q$ corrections to the Compton process
\cite{Radyushkin:2000ap} and its crossed counterpart in
$\gamma^*\gamma$ collisions \cite{Kivel:2000cz} has been completely
classified in the framework of the operator product expansion. These
corrections factorize into a hard scattering subprocess and
generalized parton distributions of twist 3, in contrast to the
twist-2 distributions discussed so far. The evolution equations of the
twist-3 distributions are known~\cite{Belitsky:2000vx}, in part to LO
and in part to NLO, whereas the NLO corrections to the hard scattering
have not been calculated as yet. It is worth mentioning that the $1/Q$
suppressed terms contribute only to amplitudes where the helicities of
the two photons differ by 1, whereas the leading contributions only
feed amplitudes where the photon helicities are equal or differ by
2. The corresponding helicity amplitudes can be separated using
appropriate angular distributions \cite{Diehl:2000uv,Diehl:1997bu};
the two-photon processes might therefore provide a window on twist-3
effects that are not masked by large twist-2 pieces.

\subsection{Meson production}

For meson electroproduction, the one-loop corrections to the hard
scattering kernels have not yet been evaluated. In the case of pion
production, they are closely connected with the one-loop corrections
to the pion form factor in the hard-scattering formalism of Brodsky
and Lepage. In fact, the Feynman diagrams for the latter can be
obtained from those for pion electroproduction (see
fig.~\ref{fig:mesons}b) by replacing the quark distribution in the
proton with the pion distribution amplitude. The NLO corrections to
the pion form factor can but need not be important, depending
crucially on the choice of renormalization scale in $\alpha_s$
\cite{Melic:1999qr}.

For the production of vector mesons, the additional calculation of the
one-loop corrections to the gluon exchange diagrams (see
fig.~\ref{fig:mesons}a) is necessary for a complete NLO evaluation. It
would be very interesting to know the size of these
corrections. Frankfurt \etal~\cite{Frankfurt:1996jw} have studied the
tree level diagrams, including in the hard scattering process the
transverse momentum $k_T$ of the quark-antiquark pair in the vector
meson, \ie, replacing the meson distribution amplitude in
fig.~\ref{fig:mesons}a with the $k_T$ dependent light-cone wave
function. This inclusion of this finite $k_T$ effects led to a very
strong suppression of the amplitude when a meson wave function was
taken that decreases as slowly as $1 /k_T^2$ at large transverse
momentum. This large-$k_T$ falloff is, however, mediated by hard gluon
exchange between the quark and antiquark forming the meson, and as
such should be included not in the meson wave function but in the hard
scattering process, where it is a \emph{part} of the NLO corrections.

\section{Summary}

Generalized parton distributions and distribution amplitudes provide
novel tools to study the interplay between partons and hadrons in
QCD. They connect several well-studied concepts such as parton
densities, distribution amplitudes, form factors, and light-cone wave
functions, and contain information beyond what can be learned from
each of these.

These novel quantities can be studied in certain exclusive processes
at large momentum transfer, whose investigation is in reach of
present-day and future experiments. The description of these processes
relies on factorization theorems and thus has a solid basis in QCD.

A quantitatively reliable extraction of generalized parton
distributions and distribution amplitudes will require a sufficient
understanding of and control over radiative corrections. The
logarithmic evolution of these quantities is well studied and the
kernels are known to NLO. As to the corrections to the hard scattering
subprocess, they are known to NLO in the case of Compton scattering
and of $\gamma^*\gamma$ collisions, but a deeper understanding of when
and why they are large is still to be achieved. Not much is known
about the NLO corrections to meson production, but some pieces of
evidence exist that they may be important in certain kinematical
situations.

\Acknowledgments 

I gratefully acknowledge the financial support of RADCOR--2000 and of
the SLAC theory group.

\end{document}